\documentclass[a4paper,12pt]{article}
\usepackage{setspace}
\def\be{\begin{equation}}
\def\ee{\end{equation}}
\def\ba{\begin{eqnarray}}
\def\ea{\end{eqnarray}}
\begin{document}

\title{{\baselineskip -.5in
\vbox{\small\hskip 4.5in \hbox{hep-th/0606218}}
\vbox{\small\hskip 4.5in \hbox{DAMTP-2006-44}}
\vbox{\small\hskip 4in \hbox{}}}
\vskip .4in
Two-charge small  black hole entropy:
  \\String-loops and 
  multi-strings} 
\author{Aninda Sinha $^1$
and Nemani V. Suryanarayana $^2$\\
{}\\
{\small{\it $^1$ Department of Applied Mathematics and Theoretical
    Physics,}}\\ 
{\small{\it Wilberforce Road, Cambridge CB3 0WA, U.K.}} \\
{\small{E-mail: {\tt A.Sinha@damtp.cam.ac.uk}}} \\
{\small{$^2$ {\it Perimeter Institute for Theoretical
      Physics,}}}\\ 
{\small{\it 31 Caroline Street North, Waterloo, ON, N2L 2Y5,
    Canada}}\\ 
{\small{E-mail: {\tt vnemani@perimeterinstitute.ca}}}
}

\maketitle \abstract{We investigate the inclusion of 10-dimensional
string loop corrections to the entropy function of two-charge extremal
small black holes of the heterotic string theory compactified on $S^1
\times T^5$ and show that the entropy is given by $\pi\sqrt{a\, q_1
q_2+b\, q_1}$ where $q_1$ and $q_2$ are the charges with $q_1 \gg q_2
\gg 1$ and $a$ and $b$ are constants. Incorporating certain
multi-string states into the microstate counting, we show that the new
statistical entropy is consistent with the macroscopic scaling for one
and two units of momentum (winding) and large winding (momentum). We
discuss our scaling from the point of view of related $AdS_3$ central
charge and counting of chiral primaries in superconformal quantum
mechanics as well.

\onehalfspacing
\vskip 1cm

\newpage
\tableofcontents
\section{Introduction}

The success of string theory in providing microscopic understanding of
the entropy of extremal black holes has been extended in recent times
to extremal small black holes with at least two charges
\cite{ad1}-\cite{senbh}. For supersymmetric two-charge black holes the
statistical entropy coming from counting the bound state degeneracy
has been checked to agree with the one calculated using Wald's entropy
formula \cite{wald} in higher-derivative supergravity
\cite{senpi,entropyfn}. Somewhat miraculously, the subleading terms
have also been shown to agree in certain cases \cite{ddmp}. The
entropy function formalism developed by Sen \cite{entropyfn} to
calculate Wald's entropy for extremal black holes has been
investigated in other contexts in \cite{ss}-\cite{others}.

The most studied extremal small black hole is the one made from a long
heterotic string wrapping a circle carrying winding charge $w$ and
momentum $n$, namely, the FP system. A simple scaling argument using
Sen's entropy function formalism \cite{senpi,entropyfn}, shows that
incorporating string tree-level higher derivative corrections gives
rise to an entropy proportional to $\sqrt{nw}$ in agreement with the
microscopic state counting. A question that naturally arises is what
happens if one includes higher 10-dimensional string-loop corrections
as well in calculating Wald's entropy. Naively, one would not
expect a simple answer to this question\footnote{{In the context of
supersymmetric black holes carrying three or more charges, this
question was partly investigated in \cite{kalloshmohaupt}}}. We will
show in this note that contrary to this naive intuition, a simple
formula for the entropy emerges when the 10-dimensional string-loop
corrections are included. The modified entropy formula predicted by
the macroscopic scaling argument will be shown to behave like
$$
S_{BH}\sim \sqrt{a\, nw+b\, n}\,,
$$
for $n\gg w\gg 1$ with $a$ and $b$ being constants. There are three
important points to note about this formula.
\begin{enumerate}
\item{When the square-root is expanded, we get corrections
  proportional to powers of $\sqrt{n/w}$ which will be more dominant
  than the subleading terms arising from the
  Hardy-Ramanujan-Rademacher formula for the entropy of the long
  string which begin with $\log (nw)$. Hence, it is important to
  understand the terms arising from the 10-dimensional string-loop
  corrections.}
\item{The formula admits a non-zero formal limit when $w \rightarrow
  0$ and we have a single charge. In this case, string-loop effects
  become very important.}
\item{Motivated by quantum entanglement arguments, Kallosh and Linde
  \cite{kalloshlinde} proposed that the single charge black hole
  entropy could be interpreted as arising from non-normalizability of
  a qubit wave-function and conjectured that the black hole entropy
  formula could be made more ``universal'' of the form $\sqrt{a q_1
  q_2+b q_1}$ for the two-charge case. Our findings nicely corroborate
  this conjecture.}
\end{enumerate}
Recently in Ref.\cite{ss} we have provided evidence for the existence
of extremal small black holes with just one electric charge with
D0-brane solution as the main example {\footnote{See also \cite{nvs}
for a discussion in AdS context.}}.
The microscopic and macroscopic entropies agreed only after including
a certain subset of non-bound states in the counting.
In general there are continuous families of classically
supersymmetric states corresponding to separating the D0-branes in the
nine spatial directions in type IIA. 
In
\cite{ss} we extracted those states from the continuous families which
have $SO(9)$ invariance (that is, no separation among D0-branes) and
associated them to the single centered D0-brane solution. This subset
of all possible non-bound states was shown to give the correct
behaviour of the leading entropy predicted by Wald's entropy
formula. This is a rather surprising result as it has been believed
that only the bound states contribute to the entropy of an extremal
black hole. Thus it raises the question of whether one should include
more than the bound state entropy for the extremal black holes with
more than one charge too.

Motivated by the above discussion of the D0-brane we will include a
subclass of the classically supersymmetric multi-string configurations
to the set of microstates of the two-charge black hole considered
above.  We again set aside the zero modes corresponding to separating
the centers of mass of the fundamental strings. By looking at two
special examples in detail, we show that this gives results for the
statistical entropy consistent with the scaling given by the entropy
function formalism.

Since the entropy function formalism depends only on the near horizon
geometry, it is expected that the statistical entropy should also
emerge from microscopic state counting in a quantum theory living at
the horizon, or equivalently on the boundary of the AdS geometry in
the spirit of AdS/CFT correspondence (see for example \cite{gsy}).  We
propose that the near horizon counting should respect the macroscopic
scaling.


In this note we outline two computations to check this
proposal. Firstly, we use the chiral primary partition function of
superconformal quantum mechanics \cite{strominger} of the D0-D4 $CY_3$
black hole to demonstrate that its asymptotics exhibit similar scaling
to our loop corrected macroscopic result for the entropy. Secondly,
motivated by the work of Kraus and Larsen \cite{kl}, we provide a
central-charge argument which can lead to the same scaling as that we
find using Sen's entropy function formalism.

The rest of this note is organized as follows. In section 2, we use
the entropy function formalism to provide a scaling argument for the
string-loop corrected small black hole macroscopic entropy. In section
3, we provide a modification of the microscopic counting which gives
rise to the same scaling. In section 4.1, we consider asymptotics of
the partition function of chiral primaries in superconformal quantum
mechanics and show that it is similar to our scaling prediction. In
section 4.2, we consider loop corrections from the central charge
viewpoint. We summarize in section 5. In the appendix, we provide a
derivation of a relation between the $AdS_3$ central charge and the
black hole entropy used in section 4.2.

\section{A scaling argument}
In this section we will explore the consequences of including the,
hitherto neglected, 10-dimensional string-loop corrections to the
entropy of the two-charge FP black hole in heterotic string compactified
on $T^6$. We use Sen's entropy function formalism \cite{entropyfn} to
calculate Wald's entropy of this black hole. The main steps in using
this formalism are as follows:
\begin{itemize}
\item{Assume a near horizon geometry of the form $AdS_2\times S^d$ and
  assume that the fields respect the isometries of the spacetime.}
\item{Write down a function $f$ defined by $f=\int \Omega_{S^d} \, L$
where $L$ is the spacetime Lagrangian.}
\item{Define electric charges $q_i$ by $q_i=\partial_{e_i} f\,,$ where
  $e_i$ are the electric fields at the horizon. Take the 
Legendre transform of $f$ only with respect to $q_i$'s (even in the
presence of magnetic fields).
$$
F=2\pi(q_i e_i-f)\,,
$$ and extremize this function with respect to the radii of $AdS_2$
and $S^d$ and scalar fields. The value of $F$ at its extremum is the
black hole entropy.}
\end{itemize}
We will assume that the near horizon geometry is going to be $AdS_2
\times S^2$ in the presence of higher derivative corrections in our
context. The 10-dimensional action for heterotic string theory in
string frame including the higher-loop corrections schematically takes
the form
\be
\int d^{10}x \,\sqrt{-G} \left[ e^{-2\phi} (R + \cdots)+
  \sum_{n=0}^\infty e^{2 n \phi} [ (n+1)- {\rm loop} ~ {\rm terms}]
  \right]\,,  
\ee
where $\cdots$ represent the $\alpha'$ corrections at the
tree-level. The string-loop terms start at the quartic order at 1-loop
and we include all the $\alpha'$ corrections at each order. This
action can be dimensionally reduced on $T^5\times S^1$ with $T$ being
the circle-size modulus and $T/S=g_s^2=e^{2\phi}$ the 10-dimensional
string coupling. This will give
\begin{equation}
\label{fullentfn}
f(v_1,v_2,v_3,v_4, S,T) = S g(v_1,v_2,v_3 = e_1T, v_4 = e_2/T) + T
h(v_1,v_2,v_3,v_4, \alpha = T/S)  
\end{equation}
Here $v_1$ and $v_2$ denote the radii of $AdS_2$ and $S^2$
respectively while $e_1$ and $e_2$ denote the two electric fields of
the Kaluza-Klein gauge fields coming from the 10-dimensional metric
and the B-field respectively. Taylor expanding the function $h(v_1,
v_2, v_3, v_4, \alpha)$ in powers of $\alpha$ accounts for all the
higher string-loop corrections in $f$. To find the near horizon
solution one has to extremize $f$ with respect to $S$, $T$, $v_1$ and
$v_2$. The $S$ and $T$ equations of motion can be written as
\begin{eqnarray}
\label{steom}
g-\alpha^2\partial_\alpha h = 0, ~~
(v_3 \partial_{v_3} - v_4 \partial_{v_4} + 1 )(g + \alpha h) = 0.
\end{eqnarray}
The two electric charges, $q_1 = T \partial_{v_3} f$, $q_2 =
\frac{1}{T} \partial_{v_4} f$ correspond to the momentum $n$ and
winding $w$ of the fundamental string. These read
\begin{eqnarray}
\label{charges1}
q_1 = ST \partial_{v_3} g + T^2 \partial_{v_3} h, ~~ q_2 = \frac{S}{T}  
\partial_{v_4} g + \partial_{v_4} h
\end{eqnarray}
which can be solved for $T$ and $\alpha$ to obtain
\begin{equation}
\label{talpha}
T^2 = \frac{q_1 \partial_{v_4} g}{(q_2 - \partial_{v_4}h)
  \partial_{v_3}g + \partial_{v_3}h \partial_{v_4}g}, ~~ \alpha =
  \frac{\partial_{v_4} g}{q_2 - \partial_{v_4} h}.
\end{equation}
Note that the right hand sides of these equations depend on $\alpha$
implicitly through $h$. The equations of motion for $v_1$ and $v_2$
are
\begin{equation}
\label{v1v2eom}
\partial_{v_i} g + \alpha \, \partial_{v_i} h = 0
\end{equation}
for $i=1$ and $2$. Then the entropy function reads
\begin{eqnarray}
S_{BH} = 2\pi (e_1 q_1 + e_2 q_2 - f) = 4 \pi \frac{v_3 q_1}{T}.
\end{eqnarray}
In obtaining the final expression we have used the equation of motion
for $T$ and the expression for $q_1$. Now we can simply substitute the
expression for $T$ from eq.(\ref{talpha}) in terms of the charges to
arrive at
\begin{equation}
\label{scale1}
S_{BH} = 4 \pi \sqrt{q_1 \, (a \, q_2 + b)}
\end{equation}
where $a=v_3^2 \partial_{v_3} g/ \partial_{v_4} g$ and $b = a \,
(\partial_{v_3} h \partial_{v_4} g - \partial_{v_4} h \partial_{v_3}
g)/\partial_{v_3} g$. However, for the coefficients $a$ and $b$ to be
universal we have to have $v_i$, $\partial_{v_i} g$ and
$\partial_{v_i} h$ appearing in $a$ and $b$ to be charge
independent. Since the eqs.(\ref{steom}) and (\ref{v1v2eom}) depend on
$\alpha$ explicitly as well as implicitly through $h$ the coefficients
in eq.(\ref{scale1}) are not completely universal. Let us first
consider two special cases where $a$ and $b$ do become universal.
\begin{itemize}
\item If we set $h=0$ then $b=0$ and the $\alpha$ dependence from
  eqs.(\ref{steom}) and (\ref{v1v2eom}) drops out. These equations
  allow us to solve for $v_i$ and hence $a$ becomes universal. In
  this case one recovers the result of Sen \cite{senpi}.
\item If we set $q_2 = 0$ then $\alpha$ and therefore $h$ become
  charge independent. In this case too the eqs.(\ref{steom}) and
  (\ref{v1v2eom}) become charge independent. Then the unknown
  coefficient $b$ in the entropy becomes universal, thus
  recovering the result of \cite{ss}.
\end{itemize}
Since we obtained eq.(\ref{fullentfn}) from the effective action of
the perturbative string, we have $\alpha \ll 1$. However since $\alpha$
can be determined through the second of eq.(\ref{charges1}) it depends
only on $q_2$. So the coefficients $a$ and $b$ in eq.(\ref{scale1})
depend, at the most, on $q_2$ but not on $q_1$. Now we would like to
argue that $a$ and $b$ are universal to the leading order in
$\alpha$. For this it is sufficient to look at the entropy function
including just the 1-loop terms. Then we have
\be 
\label{oneloopf}
f(v_1,v_2,v_3,v_4)=
\underbrace{S g(v_1,v_2,v_3,v_4)}_{{\rm
string~tree-level}}+ \underbrace{T h_0 (v_1,v_2,v_3,v_4)}_{{\rm
string~one-loop}}  \ee
where $h_0 = h(\alpha = 0)$. This leads to the entropy 
\be S_{BH}=4
\pi {q_1 v_3\over T}= 4\pi \sqrt{q_1q_2 v_3 v_4-q_1 v_3
h_0}\label{ent} 
\ee
and the near horizon values of the scalar fields
\ba
T = \sqrt{v_3 q_1 \over v_4 q_2-h_0}, ~~ S = {q_2-\partial_{v_4}
  h_0 \over \partial_{v_4}g}  \sqrt{v_3 q_1 \over v_4 q_2-h_0} \, .
\ea
Now for $\alpha = T/S \ll 1$ or in other words when $q_2\gg 1$ we can
treat the equations of motion obtained from (\ref{oneloopf})
perturbatively in $\alpha$. This results in $\partial_{v_1} g = 0, ~~
\partial_{v_2} g =0$ to the leading order. These together with $g=0$
and the perturbative approximation to the equation of motion for $T$
which is $v_3 \partial_{v_3} g= v_4 \partial_{v_4}g$ provide four
equations independent of the charges to solve for $v_1$, $v_2$, $v_3$
and $v_4$. Furthermore, note that for the higher-derivative expansion
to make sense we need $T\gg 1$ or in other words $q_1\gg
q_2$. However, it is not necessary to impose this constraint to get
the above scaling. Thus, we see that the scaling of the macroscopic
entropy is indeed of the type given by equation eq.(\ref{ent}).

One would naively have expected, based on the tree-level result $S
\sim \sqrt{q_1 q_2}$, to go to zero in the $q_2 \rightarrow 0$ limit
and hence string coupling to blow up. However $S$ and $T$ remain
sensible in the $q_2\rightarrow 0$ limit and scale the same way, i.e.,
$\sqrt{q_1}$. Thus the 10-dimensional string coupling is independent
of the charges in this limit as $g_s^2\sim T/S \sim O(1)$. This
suggests that in this limit, both $\alpha'$ and $g_s$ corrections
become equally important. 

It is the opposite limit $q_1\rightarrow 0$ that appears problematic
at first sight. This is not entirely unexpected as $T \rightarrow 0$
in this limit and hence the higher-derivative expansion breaks
down. In general, when one dimensionally reduces an action, one misses
terms in the lower dimensional theory which vanish on taking the
decompactification limit. A well-known example is the dimensional
reduction of the M-theory $R^4$ term which gives only the 1-loop type
IIA $R^4$ term while the tree-level term arises due to Kaluza-Klein
effects in the 11-dimensional theory on a circle
\cite{greenvanhove}. Such terms typically come with inverse powers of
the compact volume so that on taking the decompactification limit,
they vanish. In our context this must be the reason for the lack of
T-duality invariance of the entropy. This may be cured by making the
function $f$ invariant under T-duality $T \rightarrow 1/T$ which we
will turn to next.

\smallskip
\noindent\underline{\bf A T-dual invariant scaling:} This motivates us 
to write a T-dual invariant entropy function of the form
\begin{equation}
\label{tdualf}
f= S g+(T+{1\over T}) h_0\,,
\end{equation}
keeping terms only upto the string 1-loop order. Repeating the above
analysis for this function yields
\ba S_{BH}&=&4\pi
\sqrt{(q_1 v_3-h_0) (q_2 v_4-h_0)}\,,\\ T&=& \sqrt{q_1 v_3-h_0 \over
  q_2 v_4 -h_0}\,, \\ S&=& \sqrt{\left(q_1 q_2 v_4-q_1 h_0-(q_1
  v_3+q_2 v_4-2 h_0)\partial_{v_3}h_0\right)\left(q_1\leftrightarrow
  q_2, v_3\leftrightarrow v_4\right)\over \partial_{v_3}g
  \partial_{v_4}g (q_1 v_3-h_0)(q_2 v_4-h_0)}\,.  
\ea 
This entropy is manifestly T-duality invariant, i.e., under $q_1
\leftrightarrow q_2, v_3\leftrightarrow v_4$ as expected. In order to
argue for (leading order) universality of $v_i$ there are now three
possibilities: (a) $S\gg T\gg 1/T$ so that $n\gg w\gg 1$ or $S\gg
1/T\gg T$ so that $w\gg n\gg 1$, (b) $q_1=0$, (c) $q_2=0$. Note that
when $q_1=q_2=0$ then, $h_0=0$ and $S_{BH}=0$ as it should be. We
should however mention that the above T-duality invariant entropy
function in eq.(\ref{tdualf}) is not derived from first principles and
therefore should be considered as an illustrative toy model.

\section{Microstate re-counting}

We have seen that, when one includes string-loop corrections, the
scaling of the entropy of the two-charge extremal black hole is
different from the string tree-level one. One still needs to compute
the universal numbers $a$ and $b$ that appear in eq.(\ref{scale1})
with explicit higher derivative corrections. Unfortunately one does
not have much control over the relevant higher derivative terms. Here
we assume that $a$ and $b$ in eq.(\ref{scale1}) are going to be
non-zero.\footnote{We provide evidence in favour of this assumption in
section 5.} Such modification of the entropy cannot be accounted for
just by counting the degeneracy of the bound states, namely, the
Dabholkar-Harvey states in the first quantized string theory. This is
so because the corrections to the leading entropy coming from the
bound state counting depend only on the combination $q_1 q_2 \sim nw$
whereas string-loop corrections depend on $q_1/q_2 \sim n/w$ as well.

Motivated by the discussion of the D0-charged black hole \cite{ss} in
the introduction, we propose to include a subset of multi-strings
states which carry exactly the same quantum numbers and are (at least)
classically supersymmetric. In the case of the FP system there are
continuous zero-modes corresponding to separating the centers of mass
of different strings in a multi-string state spatially in the compact
and non-compact directions. To do a counting of discrete states we
suppress these modes and consider only one state in each such
continuous family. Then the problem boils down to counting the number
of ways one can distribute the total winding and momentum over several
different (chiral and bosonic) strings.

\subsection{Multi-string Partition function}

Let us write down a partition function including this discrete subset
of the full set of multi-string states of the FP system. Recall that
for the single string we have
\begin{equation}
m(R) = \frac{n}{R} + \frac{R \, w}{\alpha'}, ~~ N := N_L-1 = n \, w
\end{equation}
where $N_L$ is the left-moving oscillator number. Then the
single-string partition function is
\begin{equation}
Z_1(q) = \prod_{k=1}^\infty \frac{1}{(1-q^k)^{24}} = \sum_{N=0}^\infty
p_{24} (N) \, q^N.
\end{equation}
One can give a closed-form expression for the multi-string partition
function that we seek as well. The result is as follows:
\begin{eqnarray}
\label{fullpartfn}
Z(x, y) = f_1 (x) f_1(y) f_2(x, y)
\end{eqnarray}
where 
\begin{equation}
f_1 (x) = \prod_{k=1}^\infty \frac{1}{1-x^k} =
\sum_{k=0}^\infty p(n) \, x^n
\end{equation}
with $p(n)$ being the number of partitions of $n$ and
\begin{equation}
f_2 (x, y) = \prod_{k,l=1}^\infty \frac{1}{(1-
  x^k y^l)^{p_{24}(kl)}}
\end{equation}
with $p_{24}(n)$ being the number of 24-colored partitions of $n$,
generated by
\begin{equation}
\prod_{n=1}^{\infty} {1\over
  (1-x^n)^{24}}=f_{24}(x)=\sum_{l=0}^\infty p_{24}(l) \, x^l\,.
\end{equation}
To see this let us note that any multi-string state with total winding
and momentum given by $n$ and $w$ has three types of constituent
strings: (i) those carrying purely winding number, i.e, $n=0$, (ii)
those carrying purely momentum number, i.e, $w=0$ and (iii) those with
both winding and momentum non-zero. So the full partition function for
multi-string states is given by the product of three pieces. The first
two pieces are the multi-string partition functions of the first two
species of strings listed above respectively. These are the same as
the generating functions for number of partitions, namely, $f_1(x)$
and $f_1(y)$.

For the third piece, consider multi-string states with each
constituent string having momentum $k$ and winding $l$. Since the
component strings can be treated as indistinguishable bosons with each
string having $p_{24}(kl)$ `flavours', the degeneracy for a $p$-string
state is given by the number of components of a symmetric rank-$p$
tensor whose indices can take $p_{24}(kl)$ values. This number is
${(p_{24}(kl)-1+p)!\over p!  (p_{24}(kl)-1)!}$.  Therefore the
partition function for multi-string states with each string carrying
charges $(k,l)$ is:
\be 
\label{nw}
\sum_{p=0}^\infty {(p_{24}(kl)-1+p)!\over p!
(p_{24}(kl)-1)!}  (x^k y^l)^p = (1-x^k y^l)^{-p_{24}(kl)} \,.
\ee
Then the full partition function of the states with both non-zero
charges is simply given by taking the product of terms in
eq.(\ref{nw}) over non-zero $k$ and $l$ which gives $f_2 (x, y)$. This
completes the proof of the two-charge multi-string partition function
of eq.(\ref{fullpartfn}).

We should point out that there is a BPS degeneracy of 16 for each
1/4-BPS state in the single string Hilbert space coming from the
fermionic zero-modes in the right-moving sector of heterotic string
and we have not taken these into account. If we are to include them as
number of `flavours' for each string then the full partition function
will be $Z^{16}(x,y)$. It is not clear to us if one should include
this degeneracy into the counting or not. However the results below
can be modified easily to incorporate such an extra power and the
qualitative features remain the same as without these extra states.

We should next analyze{\footnote{We thank J. Lucietti for discussions
in section 3.2 and 3.3.}} the partition function in
eq.(\ref{fullpartfn}) to extract the statistical entropy. We will look
at the special cases of $n=1, 2$ and $w \gg 1$.

\subsection{Example 1: $n=1, w\gg 1$}
Consider the case when the total momentum $w\gg 1$ and winding
$n=1$. The partition function of this case is given by the coefficient
of $x$ in the full partition function which is
\begin{equation}
f_1(y)\,  [ 1+\sum_{k=1}^\infty
  p_{24}(k) \, y^k] = f_1(y) f_{24}(y) = \prod_{k=1}^\infty
  \frac{1}{(1-y^k)^{25}} = \sum_{w=0}^\infty p_{25}(w) \, y^w
\end{equation}
So the degeneracy is simply given by $p_{25}(w)$ which in the large
$w$ limit can be easily shown to yield \be \exp (\pi \sqrt{w}
\sqrt{\underbrace{16}_{ \alpha'}+\underbrace{2/3}_{\rm
string~loop}})\,.  \ee
This gives entropy $\pi \sqrt{w} \sqrt{16 + 2/3}$ consistent with that
in eq.(\ref{scale1}).

\subsection{Example 2: $n=2$, $w\gg1$}
Now consider the case $n=2$ and $w\gg1$. The full partition function
can be expanded to keep all terms of order $x^2$:
\begin{eqnarray}
&& \prod_{k=1}^\infty \frac{1}{1-y^k} \Big[ 2 + \sum_{m=1}^\infty 
    p_{24}(m) \, y^m + \frac{1}{2} \sum_{m=1}^\infty [p_{24}(m) + 
    p_{24}^2(m)] \,  y^{2m} \cr 
&& + \sum_{{m<n \atop m,n=1}}^\infty p_{24}(m) \, p_{24}(n) \, y^{m+n}
    + \sum_{n=1}^\infty p_{24}(2n) \, y^n \Big] x^2 \,.
\end{eqnarray}
Noticing that one can identify most of the terms here as those in the
expansion of
\begin{eqnarray}
 \frac{1}{2} (1+ \sum_{m=1}^\infty p_{24}(m) \, y^m) (1+
\sum_{n=1}^\infty p_{24}(n) \, y^n) 
= \frac{1}{2} + \sum_{m=1}^\infty
p_{24}(m) \, y^m + \frac{1}{2} \sum_{m,n=1}^\infty p_{24}(m) \,
p_{24}(n) \, y^{m+n}
\end{eqnarray}
and then recognizing this as simply $\frac{1}{2} (\prod_{k=1}^\infty
\frac{1}{(1-y^k)^{24}})^2 = \frac{1}{2} (\prod_{k=1}^\infty
\frac{1}{1-y^k})^{48}$ we can rewrite the above $n=2$ partition
function as
\begin{eqnarray}
\prod_{l=1}^\infty \frac{1}{1-y^l} \left[ \frac{1}{2}
  (\prod_{k=1}^\infty \frac{1}{1-y^k})^{48} + \frac{1}{2} (1+ 
\sum_{m=1}^\infty p_{24}(m) \, y^{2m}) + (1+ \sum_{n=1}^\infty
p_{24}(2n) \, y^n) \right]x^2.
\end{eqnarray}
Then the coefficient of $y^w$ reads:
\begin{equation}
\frac{1}{2} \,  p_{49}(w) + \frac{1}{2} \sum_{k=0}^w
\delta_{[\frac{w-k}{2}]-\frac{w-k}{2}} ~ p(k) \, p_{24}(\frac{w-k}{2})
+ \sum_{k=0}^w p(k) \, p_{24}(2w-2k) \,.
\end{equation}
The last term is the one coming from multi-string states with both
units of momentum carried by one single string. The 2nd term goes at
most like $p(w)$ or $p_{24}(w/2)$ and so can be dropped in comparison
to the first and the third. We will now see that the 1st and 3rd term
behave the same way for large $w$. The proof is as follows. First note
that
\be
g(x)={1\over 2}\left(f_{24}(\sqrt{x})+f_{24}(-\sqrt{x})\right)=\sum
p_{24}(2n) x^n\,. 
\ee
Then it is easy to see that the 3rd term simply becomes the
coefficient of $x^w$ in $f(x)g(x)$. Let us find out how $f(x)g(x)$
behaves near $x\rightarrow 1$ which is where we expect the maximum
contribution. $f(x)g(x)$ is made of two terms. The first term can be
written as \cite{gsw1}
\ba
\exp \left(-\sum_{n=1}^\infty \log (1-x^n)-24 \sum_{n=1}^\infty \log
(1-x^{n/2})\right )
&=&\exp\left(\sum_{m,n=1}^\infty {x^{mn}\over m}+24 \sum_{m,n=1}^\infty
   {x^{mn/2}\over m}\right) \nonumber \\
= \exp \left(\sum_m [{x^m \over m(1-x^m)}+24 {x^{m/2}\over
       m(1-x^{m/2})}]\right) \nonumber
&\sim& \exp \left ({1\over 1-x} \sum_m {1\over m^2}+{24\times 2\over
     m^2}\right) \\
&=& \exp \left({1\over 1-x} {49\pi^2\over 6}\right)\,.
\ea
The second term is
\be
\exp \left(-\sum_{n=1}^\infty \log (1-x^n)-24 \sum_{n=1}^\infty \log
(1-(-1)^n x^{n/2})\right )\,,
\ee
and following the same steps as above we get
\be
\exp\left(\sum {x^m\over m(1-x^m)}+24 {(-1)^mx^{m/2}\over
  m(1-(-1)^m x^{m/2})}\right)\,.\nonumber \\
\ee
Now writing $(-1)^m=e^{\pi i m}$ and putting $ e^{2\pi i}x=1+h$ we can
show that this term behaves exactly like the first term in
$g(x)f(x)$. These manipulations are similar to those found in
\cite{gsw1} and it would be nice to have a proof based on modular
properties of the functions also. We will not attempt this here.

Now it is easy to see that the third term will behave in the same
manner for large $w$ as the first term. Hence we conclude that for the
$n=2$, $w \gg 1$ case (similarly for $w=2$, $n \gg 1$ with $w,n$
interchanged), the entropy is given by
\be
\pi \sqrt{w} \sqrt{16\times 2+2/3}\,.
\ee
We are investigating the generalization of this formula to all $n$
separately \cite{lss}. The result to be expected for large $n,w$ may
be
$$
S_{stat}=\pi \sqrt{16 n w+{2\over 3} n+{2\over 3} w}\,,
$$ or some other formula which reduces to the cases studied above. It
will be interesting to see exactly which T-duality invariant formulae
is correct by considering the case of general $n$ and $w$ on the
counting side.

\section{Evidence from the near horizon}

Since Wald's entropy for an extremal black hole depends on the
lagrangian evaluated on the near horizon geometry alone, it is natural
to suspect that it can be evaluated by a counting of states in string
theory on the near horizon geometry of the black hole (see for example
\cite{gsy}). However we have argued so far that the entropy of the
two-charge black holes could include contributions from a subset
of non-bound states as well.

If true, these proposals should mean that the full entropy of an
extremal black hole coming from counting the corresponding BPS states
in the holographically dual superconformal quantum theory on the
boundary of the near horizon geometry should also include the
contributions coming from the subset of non-bound states like those
considered in section 3. We suggest that this is indeed the case. In
what follows we sketch two computations towards providing evidence for
our modified scaling of the entropy of eq.(\ref{scale1}). For this,
note that, even though the scaling eq.(\ref{scale1}) is derived in
heterotic on $T^6$ it should be valid for black holes in all of the
following three duality frames.
\begin{enumerate}
\item F-strings with momentum and winding in heterotic on $T^6$,
\item D0-D4 brane system in type IIA on $K3\times T^2$,
\item D1-D5 brane system in type IIB on $K3 \times T^2$.
\end{enumerate}
%
We ask whether the entropy scaling $\sqrt{n(a \, w+b)}$ could be seen
by counting the near horizon microstates in any/all of these duality
frames. Good laboratories to answer this question are the holographic
duals of string theories in the respective near horizon geometries. In
particular, the superconformal quantum mechanics of \cite{gsy} related
to the second duality frame and the 2-dimensional D1-D5 CFT relevant
for the third duality frame. Below we sketch two computations towards
extracting information of microstates which cannot be associated with
the bound states (as seen from the asymptotic flat space) from these
boundary theories.

\subsection{Superconformal quantum mechanics}

First we consider the superconformal quantum mechanics describing $D0$
branes in the $AdS_2 \times S^2 \times CY_3$ attractor geometry of a
black hole in type IIA with $D4$-branes on the $CY_3$. This quantum
mechanics has a class of chiral primaries which have been identified
with the microstates of the black hole. It was shown in \cite{gsy}
that the Bekenstein-Hawking area law for a large black hole could be
reproduced exactly from the asymptotic degeneracy of these chiral
primaries. The chiral primary generating function was argued to be
\be 
Z=\prod_{n=1}^\infty {(1+q^n)^{h_1+h_3}\over
(1-q^n)^{h_0+h_2}}=\sum_{N=0}^\infty q^N d(N)\,, 
\label{chprm} 
\ee
where 
\ba
h_0&=& D+{1\over 12} c_2\cdot P, ~~
h_1 = 3D-{3\over 4}c_2\cdot P-{\chi\over 2} \cr
h_2&=& 3D-{3\over 4}c_2\cdot P+{\chi\over 2}, ~~
h_3= D+{1\over 12} c_2\cdot P\,,
\ea
with $D=D_{ABC}P^A P^B P^C$. The $D_{ABC}$ are the intersection
numbers and $P^A$ are the D4-brane charges. The above partition
function is the same as that for a CFT with $h_1+h_3$ fermions and
$h_0+h_2$ bosons. The asymptotic value of $d(N)$ in equation
(\ref{chprm}) can be shown to be
\be
d(N)\sim \exp \left(2 \pi\sqrt{N (D-{1\over 6}c_2\cdot
P+{\chi \over 24})}\right)\,,
\ee
In \cite{gsy}, the authors only considered the leading term
proportional to $D$. However, in the two-charge small black hole
example we have to set $D=0$. Then we arrive at the relevant formula
for the statistical entropy which is clearly of the form
$\sqrt{n(w+b)}$. Furthermore, note that if we replace $CY_3$ by
$K3\times T^2$ we have $c_2\cdot P=-24 w$ with $\chi=24$. This results
in
\be
\label{predict}
S_{BH}=\pi \sqrt{16 nw+4n}\,.
\ee
Curiously, this naive replacement produces the correct leading order
term $4 \pi \sqrt{nw}$ for the two-charge black hole in type IIA on $K3
\times T^2$ with D0 and D4-branes. The entropy of this system was
worked out in a different way in \cite{kr} although the authors
obtained only the leading term in their approach.

It would be desirable to repeat the calculation for $K3\times T^2$
rigorously. If for example, one could argue convincingly that the
counting should be such that $b=0$ one would have the basis of a
non-renormalization theorem suggesting no loop corrections to the
entropy formula which would definitely merit
attention.{\footnote{Noting that the Betti numbers for $K3$,
$b_0=1,b_1=0,b_2=22,b_3=0,b_4=1$ and taking into account the Landau
degeneracy due to the magnetic field into account, our preliminary
analysis{\footnote{We thank J. Sonner for discussions on this.}}
suggests assigning an entropy of $4\pi \sqrt{n(w+1/2)}$ to the $K3
\times T^2$ small black hole. Similarly if we take $CY3$ to be $T^6$
we get $2\sqrt{2}\pi\sqrt{n(w+1)}$. Curiously, both these formulae can
be written as $\pi \sqrt{n(c_L w+c_R)}$.  From our multi-string
counting this requires replacing $f_1$ by $f_1^{c_R}$. We hope to
return to these issues in a future work.}}

We note however, that IIA on $CY_3$ is related by duality to heterotic
on $K3\times T^2$ and hence we have a prediction in eq.(\ref{predict})
for the microscopic side for this theory. The prepotential for
heterotic on $K3\times T^2$ was given, for example, in
\cite{kalloshmohaupt} and it would be very interesting to see if the
macroscopic calculation can be done in this case.

In the next subsection we pose the same question in terms of the
central charge of the CFT living on the boundary of $AdS_3 \times S^2$
which can be obtained by lifting the near horizon geometry of the
two-charge black in the third duality frame (type IIB on $K3 \times
T^2$ with D1 and D5-branes) to five dimensions.

\subsection{Corrections to $AdS_3$ central charge }

To begin with one has to establish a connection between the central
charge of $AdS_3$ and the entropy of two-charge small black holes. We
will start with the D1-D5-P system and assume that the relation
derived between the central charge and the black hole entropy
continues to hold even when one of the three charges is switched off
and we have a small black hole.

Since the central charge of $AdS_3$ is inversely proportional to
$G_3$, the 3-dimensional Newton constant and the entropy of the black
hole is inversely proportional to the 2-dimensional Newton constant
$G_2=G_3/T$, we expect the central charge $c$ of $AdS_3$ to be
proportional to $F/T$, where $F$ is the entropy function at its
extremum, i.e., the black hole entropy. It can be shown that this is
indeed the case. We get
\be
F={\pi \over 6} c(l) \tilde T\,, \label{CFT}
\ee
where $T=l^2 \tilde T$ and $F$ and $c$ are evaluated at their
extremum. Please see the appendix for a derivation of eq.(\ref{CFT}).

This relation is expected to hold for large $T$ as it involves
dimensional reduction. Chern-Simons terms can be absorbed into the
definition of $c(l)$. Although the formula (\ref{CFT}) was derived in
the appendix with the 3-charge black hole in mind we expect it to hold
for the two-charge small black hole as well. Let us cross-check this
formula (\ref{CFT}) with a known result for the two-charge small black
hole. We know that for the black hole in heterotic on $T^6$,
$F=4\pi\sqrt{nw}$, $\tilde T=\sqrt{n/w}$. Hence we expect from
(\ref{CFT}) that $c=24w$. This is precisely the central
charge{\footnote{More precisely, in presence of Chern-Simons terms the
central charge splits into $c_L$ and $c_R$ with $c_L\neq c_R$. In the
black hole entropy considered in the literature, only $c_L$ or $c_R$
enters. In the notation of \cite{kl}, this is the left-moving central
charge.}}  obtained from holographic Weyl anomaly related arguments by
Kraus and Larsen\cite{kl}.

We now expect that if there are string-loop corrections to the black
hole entropy, these can be mapped (at least when $T$ is large) to the
corrections to the central charge in $AdS_3$ associated with the black
hole. Let us give a heuristic explanation how such corrections could
arise macroscopically. It was shown in \cite{kl,saida} that the
central charge in higher derivative gravity is given by
\ba
c&=&{l_{AdS}\over 2 G_3} g_{\mu\nu} {\partial L_3\over \partial
  R_{\mu\nu}}\nonumber \\
&\sim& {l_{AdS}\over l_P}(c_0+g_s^2 c_1+g_s^4 c_2+\cdots)\,,
\ea
where $l_P\propto g_s^2$ is the Planck length and we anticipate
higher-derivative corrections in the 3-dimensional Lagrangian to be
weighted by $l_P/l_{AdS}$. Now noting that with $l_{string}=1$ we
expect for small black holes $l_{AdS}\sim O(1)$ and using Sen's
scaling result to put $g_s^2=1/w$ we find
$$
c\sim c_0 w + c_1 + c_2/w + \cdots\,,
$$ 
and hence the black hole entropy can sum to behave like
$$
S_{BH}\sim \sqrt{n (a \, w+b)}\,,
$$
as our scaling arguments predict. It will be interesting to explicitly
compute the corrections to the central charge using the technology
developed in \cite{ads3} and verify that it indeed respects the above
scaling.

\section{Discussion}

We have shown using general scaling arguments and incorporating the
10-dimensional string-loop higher derivative corrections that the
entropy of the two-charge extremal small black holes in heterotic
string on $T^6$ can get corrections that cannot be accounted for by
the standard bound state counting alone. We found that the modified
entropy behaves as
$$
S_{BH}\sim \sqrt{q_1 (a\, q_2 + b)}\,,
$$ 
where $a$ and $b$ become independent of charges in specific limits,
namely $q_1\gg q_2\gg 1$ or if one of the charges is set to zero. On
the microscopic counting side, we included by hand a certain subset of
multi-string configurations and found for two separate cases that
their counting also has the same behaviour as predicted by the scaling
argument. We presented a conjecture that for a two-charge small black
hole carrying momentum $n$, winding $w$ and $n\gg w\gg 1$, the entropy
should be given by
$$
S_{stat}=\pi \sqrt{16 nw+{2\over 3}w}\,.
$$ 
%
%
%

We further argued that Wald's entropy including string-loop
corrections can be evaluated by counting appropriate BPS states in the
near horizon geometry of the black hole. We have provided partial
evidence for this by considering the chiral primary states in the
holographically dual superconformal quantum mechanics of \cite{gsy} on
the boundary of the near horizon geometry of CY black holes in type
IIA. We also sketched a computation of string loop corrections to the
central charge of the dual CFT of $AdS_3$ and its relation to the
entropy of the two-charge black holes. It will be interesting to see
if one can recover the full partition function that we wrote down from
a counting completely within the near horizon theory or its
holographic dual. If, on the other hand, one could convincingly
demonstrate that the microstate counting dictates $b=0$ in any of the
duality frames, then it would mean that macroscopic string loop
corrections would be absent, which would merit further attention.


Finally, it will be interesting to see if there are potential
corrections to the entropy of some of the large extremal black holes
as well in string theory which may come from microstates other than
the standard bound states in analogy to the multi-string states
considered in section 3. For instance, it is possible that Wald's
entropy of the D1-D5-P black hole admits corrections so that when one
of the charges is set to zero one recovers the entropy of the
two-charge black hole that we considered here. We hope to return to
some of these questions in the near future.

\appendix

\section{Derivation of eq.(\ref{CFT})}

Here we will follow \cite{strominger} closely. The starting point is
the $AdS_3$ part of the near horizon geometry of the D1-D5-P system
which is given by
\be
ds_3^2= l^2 T^2 (dx_5+{dt\over R})^2+{U^2\over l^2} (R^2 dx_5^2
-dt^2)+l^2 {dU^2\over U^2}\,,
\ee
where the asymptotic radius of $S^1$ around which momentum $n$ flows
is given by $R$ and the variables $T,l$ are defined as
\ba
T = \sqrt{n\over k}, ~~ l^4 = g_6^2 k, ~~ k = Q_1 Q_5\,,
\ea
where $g_6$ is the six-dimensional coupling and the quantity $T$ is
the near horizon value of the radius of $S^1$. To get $AdS_2$ from
$AdS_3$ we write the metric as
\be
ds_3^2=ds_2^2+ l^2 e^{2\psi} (dx_5+ A_t dt)^2\,,
\ee
where after rescaling $t$ by $R$, $U$ by $l^2/R$ and performing the
change of variables $U^2=4 r$
\ba
e^{2\psi}&=&T^2+4r\,,\\
A_t&=& T^2 e^{-2\psi}\,,\\
ds_2^2&=&-{16\over l^2 (T^2+4r)}r^2 dt^2+l^2 {dr^2\over r^2}+l^2
e^{2\psi} (dx_5+A_t dt)^2\,. 
\ea
The $AdS_2$ factor is obtained by taking a ``near near-horizon'' limit
$U/T\rightarrow 0$. This yields after rescaling $t$ by $4/(l^2 T)$
\be
ds_3^2= l^2 (-r^2 dt^2+{dr^2\over r^2})+l^2 e^{2\psi}(dx_5+A_t
dt)^2\,, \label{nh3}
\ee
where now 
\be
A_t= {l^2\over 4} {T^3 \over T^2+4r}\approx {l^2\over 4} T(1-{4r \over
  T^2})\,,
\ee
giving $F_{rt}=-l^2/T=e$.  Following Kraus and Larsen \cite{kl}, Sahoo
and Sen \cite{sahoosen}, we define the ``central-charge'' function $c$
as follows. From equation (\ref{nh3})
\be
\sqrt{-{\rm det}G}= l^3 T\,,
\ee
then defining
\be
f_0=2\pi \sqrt{-{\rm det}G} L_0^{(3)}=-{c(l)\over 24} {T\over l^2}\,,
\ee
we get following Sen's entropy function analysis, the entropy function 
\be
F=2\pi (q e- f_0)=2\pi ({q\over \tilde T}+{c(l)\over 24} \tilde T)\,,
\ee
where $T=l^2 \tilde T$ and extremising with respect to $\tilde T,l$
yields
$$
F={\pi \over 6} c(l) \tilde T\,, 
$$
all evaluated at the extremum. This gives us our proposed relation
$c\sim F/T$ with proportionality constant $6/\pi$. 

\section*{Acknowledgements}
We thank Allen Chen, Daniel Cremades, Atish Dabholkar, Michael Green,
James Lucietti, Gautam Mandal, Samir Mathur, Rob Myers, Annamaria
Sinkovics, Julian Sonner, Ashoke Sen and David Tong for discussions.
AS acknowledges support from PPARC and Gonville and Caius college,
Cambridge. NVS would like to thank DAMTP and the Indian research
Institutes IoPB, IISc, IITM, IMSc, HRI, IACS and TIFR for hospitality
at various stages of this work.

\end{document}